\documentclass[aps,prd,onecolumn,amsmath,amssymb,superscriptaddress,preprintnumbers]{revtex4-2}

\usepackage{amsmath,amssymb,amsfonts}
\usepackage{hyperref} 
\usepackage{float}
\usepackage{mathrsfs} 
\usepackage{graphicx} 

\newcommand{\diff}{\mathrm{d}} 
\newcommand{\calO}{\mathcal{O}} 
\newcommand{\Vinv}{V_{\mathrm{inv}}} 
\newcommand{\Uinv}{U_{\mathrm{inv}}} 
\newcommand{\USch}{U_{\mathrm{Sch}}} 
\newcommand{\USdS}{U_{\mathrm{SdS}}} 
\newcommand{\kb}{\kappa_b} 
\newcommand{\kc}{\kappa_c} 

\begin{document}

\title{Bound-State Resonances of Schwarzschild-de Sitter Black Holes: Analytic Treatment}

\author{Qi-Dong Chen}
\affiliation{Department of Physics, Nanchang University, Nanchang, 330031, China}
\affiliation{Center for Relativistic Astrophysics and High Energy Physics, Nanchang University, Nanchang 330031, China}

\author{Chong-Bin Chen}
\affiliation{Department of Physics, Nanchang University, Nanchang, 330031, China}
\affiliation{Center for Relativistic Astrophysics and High Energy Physics, Nanchang University, Nanchang 330031, China}

\author{Guo-Qing Huang}
\affiliation{Department of Physics, Nanchang University, Nanchang, 330031, China}
\affiliation{Center for Relativistic Astrophysics and High Energy Physics, Nanchang University, Nanchang 330031, China}

\author{Fu-Wen Shu}
\thanks{shufuwen@ncu.edu.cn}
\affiliation{Department of Physics, Nanchang University, Nanchang, 330031, China}
\affiliation{Center for Relativistic Astrophysics and High Energy Physics, Nanchang University, Nanchang 330031, China}

\author{Tieguang Zi}
\affiliation{Department of Physics, Nanchang University, Nanchang, 330031, China}
\affiliation{Center for Relativistic Astrophysics and High Energy Physics, Nanchang University, Nanchang 330031, China}

%

\begin{abstract}
Inspired by Mashhoon’s framework connecting black hole quasi-normal modes (QNMs) to bound-state resonances in inverted potentials, V$\ddot{\text{o}}$lkel’s recent numerical analysis of asymptotically flat Schwarzschild black holes revealed a counterintuitive phenomenon: highly excited bound states rapidly delocalize, become extremely weakly bound, and exhibit wavefunctions highly sensitive to far-field perturbations. To analytically explain this phenomenon and extend the investigation to Schwarzschild–de Sitter (SdS) black holes, we derive the characteristic equation for excited bound-state resonances in SdS spacetime and obtain compact closed-form analytical expressions for their resonance energies. In the
$\Lambda\rightarrow0$ limit, our SdS-derived spectrum aligns perfectly with recent results for Schwarzschild black holes. We analytically demonstrate that the rapid and infinite delocalization of highly excited resonances is a universal feature of asymptotically flat Schwarzschild systems. 
More significantly, we prove that SdS black holes support only a finite number of bound‑state resonance levels---in sharp contrast to the infinite spectrum of the asymptotically flat case. This finiteness implies an upper bound on the oscillatory domain of the bound state resonances in SdS geometries, thereby preventing infinite delocalization and offering a fundamental distinction in the resonance structure of black holes in different asymptotic backgrounds. 
Surprisingly, we also find that delocalized half-bound states exist in SdS black holes when the $\Lambda$ takes specific discrete values. This is a unique feature of SdS black holes and is absent in asymptotically flat Schwarzschild black holes. We also reveal the deep connection between half-bound states and the number of bound-state resonance energy levels.

\end{abstract}

	
\maketitle

\section{INTRODUCTION}

Black-hole quasinormal modes (QNMs) are the characteristic damped oscillations emitted during the ringdown stage after a perturbed black hole relaxes toward equilibrium. Because their complex frequencies are determined by the background spacetime and the relevant boundary conditions, QNMs play a central role in gravitational-wave astronomy and black-hole spectroscopy \cite{dreyer2004,abbott2016prl1,abbott2016prl2,Berti:2025hly,LIGOScientific:2025wao}. They provide powerful probes of strong-field gravity and tests of general relativity \cite{berti2018,cardoso2019,franchini2024}, and they have also been discussed in connection with black-hole area quantization and related quantum-gravity questions \cite{hod1998prl}. Recent progress has further broadened the field, including new analytic treatments of QNMs spectra and the discovery of resonance phenomena such as avoided crossings and exceptional points \cite{Bolokhov:2025uxz, Wu:2026sds, Motohashi:2025qnm, Yang:2025dbn}.

For a Schwarzschild black hole, radial metric perturbations can be reduced to a master equation for the radial function $\Psi(r)$ \cite{berti2009}, which is equivalent to the time-independent Schrödinger equation
\begin{equation}
\left[ \frac{d^2}{dx^2} + \omega^2 - V(r) \right] \Psi = 0,
\label{eq:schrodinger_master}
\end{equation}
where $V(r)$ is the effective radial potential, which takes the form of a positive scattering barrier. The associated eigenvalues determine the complex QNMs frequencies.

In most physically relevant situations, the quasinormal spectrum of a black hole is not known in closed analytical form. One therefore usually relies on numerical or semi-analytical methods to compute the complex resonant frequencies \cite{press1971,delacruz1970,vishveshwara1970,davis1971,leaver1985,mashhoon1985,nollert1993,hod2003,hod2005,hod2006,hod2007,hod2013}. Although these approaches are highly successful, they do not generally yield explicit analytical relations between the spectrum and the fundamental parameters of the black-hole system. As a result, some structural features of the spectrum remain difficult to interpret directly, despite decades of progress \cite{berti2009,nollert1999,kokkotas1999,konoplya2011,Bolokhov:2025uxz}.

One particularly intriguing issue concerns the large-overtone regime. For a large class of black holes, the QNM spectrum at large overtone number differs strikingly from the oscillation spectra of more familiar classical systems \cite{kokkotas1999,nollert1993,anderson1993,motl2002,motl2003,natario2004,shu2006}. The physical origin of this behavior has long remained unclear \cite{hod1998,dreyer2003,corichi2003}. This puzzle motivated Mashhoon to propose a relation between black-hole QNMs and the bound-state resonance of the corresponding inverted potential barrier \cite{mashhoon1983}. Building on this idea, Völkel recently performed a quantitative numerical study of the bound-state spectrum and eigenfunctions associated with the inverted Schwarzschild potential \cite{volkel2025}. A key observation of that work is that highly excited states become rapidly delocalized and extremely loosely bound: their wave functions are increasingly sensitive to the far-field part of the potential, and the oscillatory region expands outward exponentially with increasing overtone number. 

At present, however, this rapid delocalization has only been identified numerically. It is therefore natural to ask whether it is merely a feature of specific numerical calculations or, instead, a generic property of the system that can be established analytically.
A major step in this direction was taken by Hod \cite{hod2025}, who derived, by purely analytical means, a closed-form expression for the infinite large-overtone bound-state resonance energy levels of the inverted potential associated with an asymptotically flat Schwarzschild black hole:
\begin{equation}
E(p,\ell;n) = -\frac{1}{16 M^2} \left\{
    \left[ \frac{\Gamma(2\ell+1)}{\Gamma(-2\ell-1)} \right]^2
    \frac{\Gamma(-\ell-p)\Gamma(-\ell+p)\Gamma(-\ell)}{\Gamma(\ell-p+1)\Gamma(\ell+p+1)\Gamma(\ell+1)}
\right\}^{\frac{1}{\ell+1/2}}
\times e^{ -\frac{2\pi n}{\sqrt{l(l+1)-1/4}} }
; \quad n \in \mathbb{Z},
\label{eq:hod_schwarzschild_energy} 
\end{equation}
where
\begin{equation}
p \equiv \sqrt{2-s^2}, \quad \ell \equiv -\frac{1}{2} + i\sqrt{l(l+1)-1/4},
\label{eq:parameter_definition:1}
\end{equation}
with $s$ denoting the spin weight of the field ($s=\{0,\pm1,\pm2\}$ for scalar, electromagnetic, and gravitational perturbations, respectively) and $l \ge s$ the angular harmonic index. Equation \eqref{eq:hod_schwarzschild_energy} provides an essential analytical tool for studying the delocalization of highly excited bound states. We note, however, that this formula does not apply to the special scalar mode with $s=0$ and $l=0$.

Hod's result applies to asymptotically flat Schwarzschild spacetimes. By contrast, the observed late-time universe is well described, to a very good approximation, by a cosmological model with positive dark-energy density, which motivates the study of black holes in asymptotically de Sitter backgrounds. In addition, another important motivation for the present work is to investigate whether the delocalization behavior derived by Hod is universal, particularly in the context of Schwarzschild--de Sitter (SdS)  black holes where the presence of a cosmological horizon alters the asymptotic boundary conditions, which may potentially influence the delocalization of highly excited bound states. In this context, the SdS spacetime provides a natural extension of the Schwarzschild geometry for investigating the impact of a positive cosmological constant on the resonance spectrum \cite{peebles2003,riess1998,spergel2003,eisenstein2005,planck2020}. The question is therefore not merely technical: it is important to understand whether the spectral properties found in the asymptotically flat case persist once $\Lambda>0$, where $\Lambda$ is the cosmological constant.

Analytical QNMs or bound-state resonance studies of SdS black holes do exist in special regimes, most notably in the near-extremal limit where the black-hole and cosmological horizons nearly coincide \cite{cardoso2003}. However, that regime is far from the parameter range relevant to the observed universe, where the two horizons are widely separated. It is therefore of clear interest to extend the recent analytical program initiated in the asymptotically flat Schwarzschild case \cite{hod2025} to the physically more realistic SdS setting.

In this paper, we develop such an extension for electromagnetic perturbations ($s=\pm1$). We derive the characteristic equation for the bound-state resonances of the inverted SdS potential and obtain the corresponding excited-state energy spectrum. We also show explicitly that, in the limit $\Lambda \to 0$, our SdS result (see Eq.~\eqref{eq:lambda_zero_energy_spectrum} below) reduces smoothly to Hod's Schwarzschild formula \eqref{eq:hod_schwarzschild_energy}. At present, analogous analytical treatments for the scalar and gravitational sectors in the SdS case are still lacking, which is why we focus here on the electromagnetic mode.

We then use analytical method to the delocalization problem. For the asymptotically flat Schwarzschild black hole, we prove that the rapid outward expansion of the oscillatory region observed numerically in Ref.~\cite{volkel2025} is not a numerical artifact, but a generic feature of highly excited bound-state resonances for all modes satisfying $l>0$, $s\in\{0,\pm1,\pm2\}$, and $l\ge |s|$. In particular, we derive an explicit expression for the exponentially fast growth of the oscillatory domain with increasing principal quantum number $n$ (see Eq.~\eqref{eq:63} below). Because the asymptotically flat Schwarzschild system supports an infinite tower of bound-state resonances, this delocalization is unbounded.

However, we show that the situation changes qualitatively for the SdS black hole. In the presence of a positive cosmological constant, the exponentially rapid delocalization that characterizes the asymptotically flat Schwarzschild case is absent (see Eq.~\eqref{eq:75} below). More importantly, we prove that the SdS system admits only a finite number of bound-state resonance levels, in sharp contrast with the infinite spectrum of the asymptotically flat case. Consequently, the oscillatory region of the corresponding eigenfunctions is bounded from above, and unbounded delocalization cannot occur.

Intriguingly, we uncover a peculiar half-bound state residing in the inverted potential of SdS spacetimes: the zero-energy resonance. This half-bound state is delocalized. Moreover, it arises solely in SdS black holes for specific discrete values of the cosmological constant, and is entirely absent in the asymptotically flat Schwarzschild limit. We present a detailed investigation of the distinctive properties of this half-bound state and elucidate its profound physical implications.

The paper is organized as follows.  In Sec. II, we give a detailed description of the bound-state resonance system of the SdS black hole.  In Sec. III, we derive the characteristic equation governing the bound-state resonances. In Sec. IV, we derive the analytical expression for the energy levels of bound-state resonances in the limit $\Lambda \to 0$, and compare our results with Hod's calculations\eqref{eq:hod_schwarzschild_energy} to verify their consistency. In Sec.  V, we derive the analytical expressions for the energy spectrum of bound-state resonances of the SdS black hole within specific parameter regime and investigate the extent to which the $\Lambda$ affects the resonance energy levels. In Sec. VI, we use analytical approaches to separately investigate the delocalization phenomenon of high energy levels bound-state resonances in the excited states of the asymptotically flat Schwarzschild black hole and the SdS black hole, and further analyze the discrepancies between the two cases. In Sec. VII, we present the delocalized half-bound states unique to SdS black holes and explore their distinctive properties and physical implications. In Section VIII, we summarize the main conclusions of this work and present an outlook on future research.

Throughout this paper, we adopt the natural units with $c=G=1$.

\section{BOUND-STATE RESONANCE SYSTEM OF THE SDS BLACK HOLE}
The curved spacetime line element of an SdS black hole is given by
\begin{equation}
ds^2 = f(r) dt^2 - \frac{dr^2}{f(r)} - r^2 d\Omega^2, \quad f(r) = 1 - \frac{2M}{r} - \frac{\Lambda r^2}{3}.
\label{eq:sds_metric}
\end{equation}
This system features dual horizons, namely the black hole horizon and the de Sitter horizon. The black hole horizon radius is denoted as $r_h$, and the de Sitter horizon radius as $r_c$, which satisfy
\begin{equation}
f(r_h) = 0, \quad f(r_c) = 0.
\label{eq:horizon_condition}
\end{equation}

The dynamics of the linearized field mode $\Psi$ can be radially written in the form of a Schrödinger equation \cite{berti2009}
\begin{equation}
\left[ \frac{d^2}{dx^2} + \omega^2 - V(r;M,s,l) \right] \Psi = 0,
\label{eq:sds_schrodinger}
\end{equation}
where the ``tortoise'' radial coordinate $x$ is defined by the following differential relationship
\begin{equation}
\frac{dx}{dr} = \frac{1}{f(r)},
\label{eq:tortoise_coord}
\end{equation}
and the scattering potential $V$ is given by \cite{berti2009}
\begin{equation}
V(r;M,s,l) = \left( 1 - \frac{2M}{r} - \frac{\Lambda r^2}{3} \right) \left[ \frac{l(l+1)}{r^2} + (1-s^2)\left( \frac{2M}{r^3} - \frac{(4-s^2)\Lambda}{6} \right) \right].
\label{eq:general_scattering_potential}
\end{equation}
The inverted SdS potential is as follows \cite{volkel2025,mashhoon1983,hod2025}
\begin{equation}
V_{\text{inv}}(r;M,s,l) = -V(r;M,s,l) .
\label{eq:define inverted potential}
\end{equation}
Under electromagnetic perturbations ($s=\pm1$), the scattering potential barrier is
\begin{equation}
V(r;M,s,l) = \left( 1 - \frac{2M}{r} - \frac{\Lambda r^2}{3} \right) \frac{l(l+1)}{r^2},
\label{eq:em_scattering_potential}
\end{equation}
and the inverted SdS potential for $s=\pm1$ can be written in the following form 

\begin{equation}
V_{\text{inv}}(r;M,s,l) = - \left( 1 - \frac{2M}{r} - \frac{\Lambda r^2}{3} \right) \frac{l(l+1)}{r^2}.
\label{eq:inverted_sds_potential}
\end{equation}

Our objective  is to analytically calculate the expression for the energy levels spectrum that characterize the bound-state resonances of the inverted SdS black-hole potential under electromagnetic perturbations ($s=\pm1$). When $s=\pm1$ the definition of the effective spin index and harmonic index of the field \eqref {eq:parameter_definition:1} can be written in the following form 
\begin{equation}
p \equiv |s| = 1, \quad \ell \equiv -\frac{1}{2} + i\sqrt{l(l+1)-1/4}.
\label{eq:parameter_definition}
\end{equation}
The combination of the Schrödinger equation \eqref{eq:sds_schrodinger} and the binding potential \eqref{eq:inverted_sds_potential} can be written in the following form:
\begin{equation}
\left\{ \frac{d^2}{dx^2} + E - \left( 1 - \frac{2M}{r} - \frac{\Lambda r^2}{3} \right) \frac{\ell(\ell+1)}{r^2} \right\} \Psi = 0,
\label{eq:schrodinger_inverted_potential}
\end{equation}
where $E$ is the bound-state resonance energy of the system and is defined by
\begin{equation}
E = \omega^2 ,
\label{eq:bound_state_energy_def}
\end{equation}
with the bound-state resonance energy $E\in\mathbb{R}$, $i\omega\in\mathbb{R}$, and the eigenfunction $\Psi$ is bounded:
\begin{equation}
\Psi(x\to\pm\infty) \to 0.
\label{eq:boundary_condition}
\end{equation}

\section{CHARACTERISTIC EQUATION FOR BOUND-STATE RESONANCE OF THE INVERTED SdS POTENTIAL}
In this section, we develop an analytical technique to derive the characteristic equation for bound-state resonances of the inverted SdS potential for electromagnetic perturbation ($s=\pm1$), which can be used to solve for the frequency spectrum $\{\omega_n\}$ and energy level spectrum $\{E_n\}$ of the system at large-$n$ (excited) energy levels ($n\gg1$). Recent observations show that the characteristic gravitational length of the known most massive black hole  $M\ll\Lambda^{-1/2}$ \cite{planck2020,mcdonald2012}. It is therefore reasonable to assume that the SdS black-hole system satisfies the dimensionless inequality
\begin{equation}
M\sqrt{\Lambda} \ll 1,
\label{eq:small_lambda_inequality}
\end{equation}
meaning that there is a vast distance between the de Sitter horizon $r_c$ and the black hole horizon $r_h$ ($r_h\ll r_c$).

On the other hand, the numerical results presented in Ref.\cite{volkel2025} show that the large-$n$ excited-state energy levels ($n\gg1$) of Schwarzschild black holes in an asymptotically flat spacetime background ($\Lambda=0$) satisfy the following dimensionless inequalities
\begin{equation}
M|\omega_n| \ll 1, \quad M^2|E_n| \ll 1 \quad \text{for } n\gg1.
\label{eq:small_energy_inequality}
\end{equation}
The large-$n$ bound-state energy levels of SdS black holes are approximately consistent with those of asymptotically flat Schwarzschild black holes in the small-$\Lambda$ regime, because the cosmological constant correction term $\Lambda$ affects the bound-state resonance energy levels as a perturbation term in this regime, which will be proven in the following. That is to say, the large-$n$ excited-state energy levels of SdS black holes analyzed in this paper also satisfy the dimensionless inequality \eqref{eq:small_energy_inequality}.

In what follows, we will perform a matching analysis on the ordinary differential equation within the regime defined by Eqs. \eqref{eq:small_lambda_inequality} and \eqref{eq:small_energy_inequality} to analytically derive the characteristic equation for bound-state resonances of the inverted SdS potential \eqref{eq:inverted_sds_potential}.

First, the ordinary differential equation \eqref{eq:schrodinger_inverted_potential} can be equivalently written in the form of the radial Teukolsky equation 
\cite{teukolsky1973}
\begin{equation}
\Delta_r^{-p+1} \frac{d}{dr} \left( \Delta_r^{p+1} \frac{d\Psi}{dr} \right) + \left[ \omega^2 r^4 + 2i\omega p r^3 -6iM\omega p r^2 + \Delta_r \left( \frac{2}{3}\Lambda(p+1)(2p+1)r^2 + 2p - \lambda \right) \right] \Psi = 0,
\label{eq:teukolsky_equation}
\end{equation}
where
\begin{equation}
\lambda = \ell(\ell+1) - p(p-1),
\label{eq:lambda_def}
\end{equation}
and the definition of $\Delta_r$ is
\begin{equation}
\Delta_r \equiv r^2 f(r) = r^2 - 2Mr - \frac{\Lambda r^4}{3}.
\label{eq:delta_r_def}
\end{equation}
The effective field parameters $p$ and $\ell$ are given by Eq. \eqref{eq:parameter_definition}.

Equation \eqref{eq:teukolsky_equation} is the Teukolsky equation for a black hole with zero spin, which is physically equivalent to Eq. \eqref{eq:schrodinger_inverted_potential} when the black hole spin vanishes \cite{chandrasekhar1975}. We focus on the bound-state resonance system, which is characterized by spatially regular (bounded) radial eigenfunctions satisfying Eq. \eqref{eq:boundary_condition}, with the asymptotic properties
\begin{equation}
\Psi(x\to-\infty) \sim e^{|\omega|x} \to 0,
\label{eq:near_horizon_asymptotic}
\end{equation}
and
\begin{equation}
\Psi(x\to+\infty) \sim e^{-|\omega|x} \to 0.
\label{eq:far_horizon_asymptotic}
\end{equation}
In the following, we analytically derive the characteristic equation for bound-state resonances of the inverted SdS potential via the asymptotic matching technique within the dimensionless regime defined by Eqs. \eqref{eq:small_lambda_inequality} and \eqref{eq:small_energy_inequality}.

\subsection{Near black-hole horizon region}
We define the near black-hole horizon region $r\sim r_h$ as
\begin{equation}
r \ll r_c, \quad |\omega|r \ll 1,
\label{eq:near_horizon_region_def1}
\end{equation}
i.e.
\begin{equation}
r\sim r_h \ll \min\left( \sqrt{3/\Lambda}, 1/|\omega| \right).
\label{eq:near_horizon_region_def2}
\end{equation}
In this region, the terms containing $\Lambda$ in Eq. \eqref{eq:schrodinger_inverted_potential} can be neglected as infinitesimal terms. Under this approximation, Eq. \eqref{eq:schrodinger_inverted_potential} is approximately equivalent to
\begin{equation}
\left\{ \frac{d^2}{dx^2} + \omega^2 - \left( 1 - \frac{2M}{r} \right) \frac{\ell(\ell+1)}{r^2} \right\} \Psi = 0,
\label{eq:near_horizon_approx_schrodinger}
\end{equation}
Similarly, the terms containing $\Lambda$ in the Teukolsky equation \eqref{eq:teukolsky_equation} are neglected as infinitesimal terms in this region. That is to say, the SdS black hole system is approximately equivalent to a Schwarzschild black hole system in an asymptotically flat spacetime background in this region \cite{hod2025}. To this end, we define the dimensionless variables
\begin{equation}
z \equiv \frac{r-2M}{2M}, \quad k \equiv 2M|\omega|,
\label{eq:near_horizon_dimless_var}
\end{equation}
such that the radial differential equation can be written in the following form \cite{hod2025}
\begin{multline}
z^2(z+1)^2 \frac{d^2\Psi}{dz^2} + (p+1)z(z+1)(2z+1) \frac{d\Psi}{dz} + \left[ -k^2 z^4 - 2pk z^3 - \ell(\ell+1)z(z+1) + pk(2z+1) - k^2 \right] \Psi = 0.
\label{eq:near_horizon_z_eq}
\end{multline}

In the radial region satisfying the boundary conditions, the mathematical solution of the differential equation is given by the following functional expression \cite{hod2025,abramowitz1970}
\begin{equation}
\Psi(z) = z^{-p+k} (z+1)^{-p-k} \,_2F_1(-\ell-p, \ell-p+1; 1-p+2k; -z),
\label{eq:near_horizon_solution}
\end{equation}
where $_2F_1(a,b;c;z)$ is the hypergeometric function \cite{abramowitz1970}.

\subsection{Near de Sitter horizon region}
We define the near de Sitter horizon region as
\begin{equation}
r\sim r_c \gg r_h,
\label{eq:far_region_def1}
\end{equation}
or equivalently
\begin{equation}
r\sim r_c \gg M.
\label{eq:far_region_def2}
\end{equation}
In this region, the terms containing the mass $M$ in Eq. \eqref{eq:schrodinger_inverted_potential} can be neglected as infinitesimal terms, and equation \eqref{eq:schrodinger_inverted_potential} can be approximately equivalent to
\begin{equation}
\left\{ \frac{d^2}{dx^2} + \omega^2 - \left( 1 - \frac{\Lambda r^2}{3} \right) \frac{\ell(\ell+1)}{r^2} \right\} \Psi = 0.
\label{eq:far_region_approx_schrodinger}
\end{equation}
The SdS black hole system in this region is approximately equivalent to a massless spin field system in a de Sitter spacetime background \cite{suzuki1996}.

We again define the dimensionless variables
\begin{equation}
q \equiv r/a, \quad y \equiv \frac{1-q}{1+q},
\label{eq:far_region_dimless_var1}
\end{equation}
where
\begin{equation}
a = \sqrt{\frac{3}{\Lambda}}.
\label{eq:far_region_dimless_var2}
\end{equation}
The radial equation \eqref{eq:teukolsky_equation} is then expressed as \cite{suzuki1996}
\begin{multline}
 \frac{d^2}{dy^2}  \Psi + \left[ (p+1)\frac{1}{y} - 2(p+1)\frac{1}{1-y} - 2(2p+1)\frac{1}{y+1} \right] \frac{d}{dy}\Psi \\
- \Bigg\{ (\ell-p)(\ell+p+1)\frac{1}{(1-y)^2} - (2p+1)(2p+2)\frac{1}{(y+1)^2} + \left[ \frac{(a|\omega|)^2}{4} - \frac{a|\omega| p}{2} \right] \\
+ \left[ -(\ell-p)(\ell+p+1) +a|\omega|p \right]\frac{1}{y(y-1)} + (2p+1)(2p+2)\frac{1}{y(y+1)} \Bigg\} \Psi = 0.
\label{eq:far_region_y_eq}
\end{multline}

In the radial region satisfying the physically motivated boundary conditions, the mathematical solution of the differential equation is given by the functional expression \cite{abramowitz1970,suzuki1996}
\begin{equation}
\Psi(y) = y^{\frac{a|\omega|}{2}-p} (1-y)^{\ell-p} (y+1)^{2p+1} \,_2F_1(\ell+1-p,\ell+1+a|\omega|;1-p+a|\omega|;y).
\label{eq:far_region_solution}
\end{equation}

\subsection{Matching of Solutions in the radial overlap region}
The radial overlap region is
\begin{equation}
M \ll r \ll \min\left( \sqrt{3/\Lambda}, 1/|\omega| \right).
\label{eq:overlap_region_def}
\end{equation}
Eq. \eqref{eq:near_horizon_solution} containing the hypergeometric function can be expanded in the region \eqref{eq:near_horizon_region_def2} into the following form \cite{abramowitz1970}
\begin{equation}
\Psi_{\text{near}}(r) \simeq A \left( \frac{r}{2M} \right)^{\ell-p} + B \left( \frac{r}{2M} \right)^{-\ell-p-1},
\label{eq:near_horizon_expansion}
\end{equation}
where
\begin{equation}
A = \frac{\Gamma(1-p+2k)\Gamma(2\ell+1)}{\Gamma(\ell-p+1)\Gamma(\ell+1+2k)}, \quad B = \frac{\Gamma(1-p+2k)\Gamma(-2\ell-1)}{\Gamma(-\ell-p)\Gamma(-\ell+2k)}.
\label{eq:near_horizon_coeffs}
\end{equation}

Similarly, Eq. \eqref{eq:far_region_solution} containing the hypergeometric function can be expanded in the region \eqref{eq:far_region_def2} into the following form \cite{abramowitz1970}
\begin{equation}
\Psi_{\text{far}}(r) \simeq \alpha \, r^{\ell-p} + \beta \, r^{-\ell-p-1},
\label{eq:far_region_expansion}
\end{equation}
where
\begin{equation}
\alpha = 2^{\ell+p+1} a^{-\ell+p} C_1, \quad \beta = 2^{-\ell+p} a^{\ell+p+1} C_2,
\label{eq:far_region_alpha_beta}
\end{equation}
and
\begin{equation}
C_1 = \frac{\Gamma(1-p+a|\omega|)\,\Gamma(-2\ell-1)}{\Gamma(-\ell+a|\omega|)\,\Gamma(-\ell-p)}, \quad C_2 = \frac{\Gamma(1-p+a|\omega|)\,\Gamma(2\ell+1)}{\Gamma(\ell+1-p)\,\Gamma(\ell+1+a|\omega|)}.
\label{eq:far_region_C_coeffs}
\end{equation}

Next, we match the coefficients of the two solutions in the radial overlap region \eqref{eq:overlap_region_def}, yielding
\begin{equation}
\frac{B}{A} \, (2M)^{2\ell+1} = \frac{\beta}{\alpha}.
\label{eq:matching_condition}
\end{equation}
Substituting Eqs. \eqref{eq:near_horizon_coeffs}, \eqref{eq:far_region_alpha_beta} and \eqref{eq:far_region_C_coeffs}, we obtain the characteristic equation for the angular frequency $\omega$:
\begin{equation}
(4M)^{2\ell+1} \left[ \frac{\Gamma(-2\ell-1)}{\Gamma(2\ell+1)} \right]^2 \left[ \frac{\Gamma(\ell-p+1)}{\Gamma(-\ell-p)} \right]^2 \frac{\Gamma(\ell+1+2k)}{\Gamma(-\ell+2k)} = a^{2\ell+1} \frac{\Gamma(-\ell+a|\omega|)}{\Gamma(\ell+1+a|\omega|)}.
\label{eq:sds_characteristic_eq}
\end{equation}
Eq. \eqref{eq:sds_characteristic_eq} is the characteristic equation for bound-state resonances of the inverted SdS potential, which is one of the core conclusions of this paper.

\section{VERIFICATION IN THE LIMIT OF $\Lambda \to 0^+$}
In this section, we calculate the asymptotic behavior of the solution to Eq. \eqref{eq:sds_characteristic_eq} in the limit where the cosmological constant term $\Lambda\to0^+$, and compare it with the calculation results of Hod \eqref{eq:hod_schwarzschild_energy} to verify their consistency.

We define $k^{(0)}$, $E^{(0)}$ as the limit of $k$, $E$ in the $\Lambda \to 0^+$ regime
\begin{equation}
\Lambda\to0^+, \quad k\to k^{(0)}, \quad E\to E^{(0)}.
\label{eq:lambda_zero_limit_def}
\end{equation}
In this limit $a|\omega|\to +\infty$, and from Eq. \eqref{eq:sds_characteristic_eq} we can obtain the resonance equation (see Appendix \ref{app:gamma_asymptotic} for details)
\begin{equation}
(2k^{(0)})^{2\ell+1} = \left[ \frac{\Gamma(2\ell+1)}{\Gamma(-2\ell-1)} \right]^2 \frac{\Gamma(-\ell-p)^2 \Gamma(-\ell+2k^{(0)})}{\Gamma(\ell-p+1)^2 \Gamma(\ell+1+2k^{(0)})}.
\label{eq:lambda_zero_resonance_eq}
\end{equation}
The discrete solutions in the small-$k$ regime (see Eqs. \eqref{eq:small_energy_inequality} and \eqref{eq:near_horizon_dimless_var}) are given by the dimensionless functional expression
\begin{equation}
k^{(0)}(p,\ell;n) = \frac{1}{2} \left\{ \left[ \frac{\Gamma(2\ell+1)}{\Gamma(-2\ell-1)} \right]^2 \frac{\Gamma(-\ell-p)^2 \Gamma(-\ell)}{\Gamma(\ell-p+1)^2 \Gamma(\ell+1)} \right\}^{\frac{1}{2\ell+1}} \times e^{-\frac{i\pi n}{\ell+1/2}}, \quad n\in\mathbb{Z}.
\label{eq:k0_solution}
\end{equation}
Combining with Eq. \eqref{eq:bound_state_energy_def}, we obtain the expression for the bound-state resonance energy levels spectrum
\begin{equation}
E^{(0)}(p,\ell;n) = -\frac{1}{16M^2} \left\{ \left[ \frac{\Gamma(2\ell+1)}{\Gamma(-2\ell-1)} \right]^2 \frac{\Gamma(-\ell-p)^2 \Gamma(-\ell)}{\Gamma(\ell-p+1)^2 \Gamma(\ell+1)} \right\}^{\frac{1}{\ell+1/2}} \times e^{-\frac{2\pi n}{\sqrt{l(l+1)-1/4}}}, \quad n\in\mathbb{Z}.
\label{eq:lambda_zero_energy_spectrum}
\end{equation}

Next, we will verify the consistency between Eq.\eqref{eq:lambda_zero_energy_spectrum} and Eq.\eqref{eq:hod_schwarzschild_energy}. From the properties of the Gamma function \cite{abramowitz1970}, we have (see Appendix \ref{app:gamma_recurrence} for details)
\begin{equation}
\frac{\Gamma(-\ell-1)\Gamma(-\ell+1)}{\Gamma(\ell)\Gamma(\ell+2)} = \frac{\Gamma(-\ell-1)\, \ell(\ell+1)\Gamma(-\ell-1)}{\Gamma(\ell)\, \ell(\ell+1)\Gamma(\ell)} = \frac{\Gamma(-\ell-1)^2}{\Gamma(\ell)^2}.
\label{eq:gamma_identity}
\end{equation}
Taking cognizance of Eqs. \eqref{eq:parameter_definition:1},\eqref{eq:parameter_definition} and \eqref{eq:gamma_identity}, we can obtain that Eq. \eqref{eq:lambda_zero_energy_spectrum} is exactly equivalent to Eq. \eqref{eq:hod_schwarzschild_energy} when $s=\pm1$. In other words, in the limit where the cosmological constant $\Lambda\to0^+$, the bound-state resonance energy level spectrum \eqref{eq:lambda_zero_energy_spectrum} obtained by solving the characteristic equation \eqref{eq:sds_characteristic_eq} is fully consistent with the results calculated by Hod (Eq. \eqref{eq:hod_schwarzschild_energy}).

\section{RESONANCE ENERGY LEVELS OF SDS BLACK HOLE}
In this section we would like to obtain the resonance energy levels of SdS black hole by solving the characteristic equation \eqref{eq:sds_characteristic_eq} for electromagnetic perturbation modes ($s=\pm1$). It turns out that analytical formulas for the excited bound-state resonance energy levels $(n\gg1)$ of SdS black hole can be obtained in specific spectral interval.

Expanding Eq. \eqref{eq:sds_characteristic_eq} in the small-$k$, large-$a$ regime (see Eqs. \eqref{eq:small_lambda_inequality}, \eqref{eq:small_energy_inequality}, \eqref{eq:near_horizon_dimless_var} and \eqref{eq:far_region_dimless_var2}) to the order that explicitly includes the $\Lambda$ correction term, we obtain
\begin{eqnarray}
&&(4M)^{2\ell+1} \left[ \frac{\Gamma(-2\ell-1)}{\Gamma(2\ell+1)} \right]^2 \left[ \frac{\Gamma(\ell-p+1)}{\Gamma(-\ell-p)} \right]^2 \frac{\Gamma(\ell+1+2k)}{\Gamma(-\ell+2k)} \nonumber\\
&=& a^{2\ell+1} (a|\omega|)^{-2\ell-1} \left\{ 1 + \frac{\ell(\ell+1)(2\ell+1)}{6(a|\omega|)^2} + \calO\left[ (a|\omega|)^{-3} \right] \right\}.
\label{eq:characteristic_eq_expanded}
\end{eqnarray}
Rearranging the equation, we obtain
\begin{equation}
(2k)^{2\ell+1} = \left[ \frac{\Gamma(2\ell+1)}{\Gamma(-2\ell-1)} \right]^2 \frac{\Gamma(-\ell-p)^2 \Gamma(-\ell+2k)}{\Gamma(\ell-p+1)^2 \Gamma(\ell+1+2k)} \left\{ 1 + \frac{2\ell(\ell+1)(2\ell+1) M^2 \Lambda}{9k^2} + \calO\left[(a|\omega|)^{-3} \right] \right\}.
\label{eq:k_characteristic_eq}
\end{equation}

Next, let us make the following definition
\begin{equation}
k \equiv k^{(0)} \left(1+\delta\right).
\label{eq:49}
\end{equation}
 In the spectral interval
\begin{equation}
 M\sqrt{\Lambda} \ll  k^{(0)}(p,\ell;n)\ll1;\quad n \in \mathbb{Z},
	\label{interval:1}
\end{equation}
the correction term $|\delta| \ll 1$.
Combining Eqs.~\eqref{eq:lambda_zero_resonance_eq}, \eqref {eq:k_characteristic_eq} and \eqref{eq:49}, we obtain the correction term $\delta$ (truncate at leading order) as
\begin{equation}
\delta =\frac{2\ell(\ell+1) M^2 \Lambda}{9 \left[k^{(0)}(p,\ell;n)\right]^2 } ;\quad n \in \mathbb{Z},
\label{eq:51}
\end{equation}
and we obtain the resonance equation by Combining Eqs.~\eqref{eq:49} and \eqref{eq:51}
\begin{equation}
	k(p,\ell;n) = k^{(0)}(p,\ell;n) \left[ 1 + \frac{2\ell(\ell+1) M^2 \Lambda}{9 \left[k^{(0)}(p,\ell;n)\right]^2 } \right]; \quad n \in \mathbb{Z}.
	\label{k:{1}}
\end{equation}
Substituting into Eq.~\eqref{eq:bound_state_energy_def}, we obtain the expression for the energy levels 
\begin{equation}
	E(p,\ell;n) = E^{(0)}(p,\ell;n)  \left[ 1 - \frac{\ell(\ell+1) \Lambda}{18 E^{(0)}(p,\ell;n) } \right] ^2; \quad n \in \mathbb{Z}.
	\label{E:{1}}
\end{equation}

The $\Lambda$-induced relative shift $\xi_{n}$ of the bound-state resonance energy spectrum is a dimensionless physical quantity characterizing the magnitude of the correction to the bound-state resonance energy levels by $\Lambda$, defined as the relative deviation between the resonance energy level in the presence of $\Lambda$ and the reference energy level in the absence of $\Lambda$, where
\begin{equation}
\xi_{n} =\frac{E(p,\ell;n)-E^{(0)}(p,\ell;n)}{E^{(0)}(p,\ell;n)};\quad n \in \mathbb{Z}.
\label{eq:xi}
\end{equation}
Combining Eqs.\eqref{eq:xi},\eqref{eq:lambda_zero_energy_spectrum}and \eqref{E:{1}}, we obtain the curves of the relative shift 
\begin{equation}
\xi_{n} =\frac{-16l(l+1) M^2\Lambda}{ 9 \left[\frac{\Gamma(2\ell+1)^2\Gamma(-\ell-p)^2 \Gamma(-\ell)}{\Gamma(-2\ell-1)^2\Gamma(\ell-p+1)^2 \Gamma(\ell+1)} \right]^{\frac{1}{\ell+1/2}}}  \times e^{\frac{2\pi n}{\sqrt{l(l+1)-1/4}}}; \quad n \in \mathbb{Z}.
	\label{eq:xi_n}
\end{equation}

Evidently, within spectral interval \eqref{interval:1}, the relative shift grows exponentially with increasing $n$. For the even higher energy level  $k^{(0)}(p,\ell;n)\sim\calO{(M\sqrt{\Lambda})} $ , the correction term $\delta\sim\calO{(1)}$ is no longer a small quantity. From the above analysis, we can draw the conclusion that the resonance energy levels of higher excited states are more sensitive to the correction term of the cosmological constant $\Lambda$ , and even an arbitrarily small $\Lambda$ will exert a non-negligible effect on specific spectral intervals of high-lying excited states.

\section{Delocalization of Bound-State Resonances}

The analytical expressions for the energy spectrum of bound-state resonances in the inverted potential provide a critical tool for analytically explaining the origin of the rapid delocalization of highly excited bound-states of the inverted potential well of asymptotically flat Schwarzschild black holes, as observed via numerical simulations in Ref.~\cite{volkel2025}. In this section, we demonstrate, analytically, how this is achieved, and discuss the corresponding phenomena in the bound-state resonance system of SdS black holes.

To proceed, we notice that the region satisfying the following inequality is called the classically allowed region for bound-state resonances. This is also the oscillatory region of the bound-state resonance eigenfunctions:
\begin{equation}
\Vinv(r; M, s, l) < E_n,
\label{eq:56}
\end{equation}
where $\Vinv(r; M, s, l)$ is the inverted Regge–Wheeler potential, the expression for $\Vinv(r; M, s, l)$ is given in Eq.~\eqref{eq:define inverted potential}. 

\begin{figure}[H]
    \centering
    \includegraphics [width=0.6\linewidth]{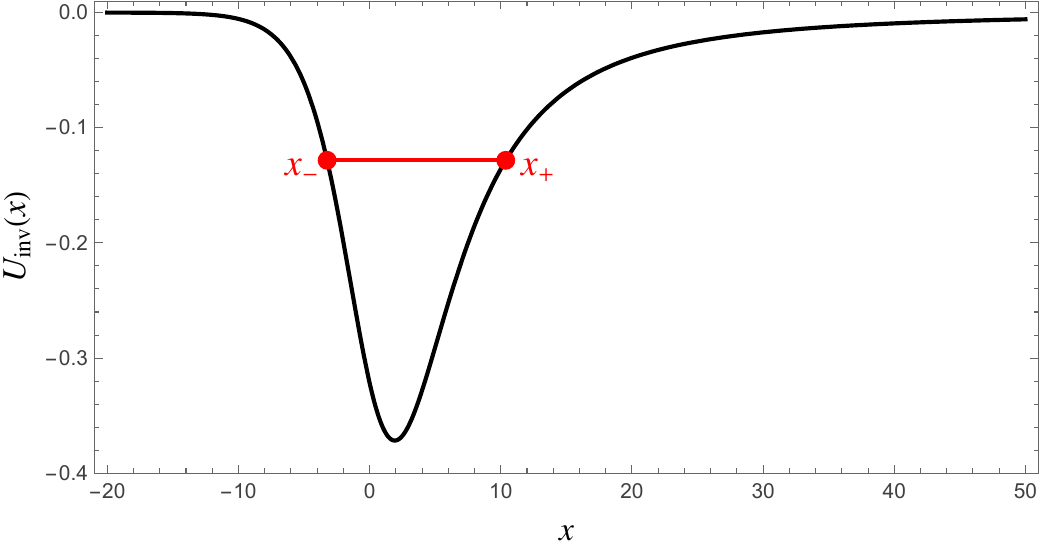}
    \caption{This is a schematic diagram of the inverted potential function (black curve) and one of its bound-state resonance energy levels  $E_n$  (red line segment). The region between the classical turning points  $x_-$  and  $x_+$ is the classically allowed region (oscillatory region) of this bound-state resonance, and the regions where $x<x_-$ and $x>x_+$ are its classically forbidden regions (evanescent regions).}
    \label{fig:placeholder}
\end{figure}

In terms of the tortoise coordinate, the extent of the classically allowed region for bound-state resonances is defined as
\begin{equation}
x_- < x < x_+,
\label{eq:57}
\end{equation}
where $x_+$ and $x_-$ are the classical turning points at the two boundaries of the classically allowed region (see Fig. \ref{fig:placeholder}), satisfying
\begin{equation}
\Uinv(x_-; M, s, l) =  \Uinv(x_+; M, s, l) = E_n,
\label{eq:58}
\end{equation}
$\Uinv(x; M, s, l)$ denotes the inverted Regge–Wheeler potential in the tortoise coordinate.

The eigenfunctions of bound-state resonances oscillate in the classically allowed region, and decay rapidly in the classically forbidden region. If the width of the classically allowed region ($x_+-x_-$) is finite, the bound-state resonance is localized. Conversely, if the width of the classically allowed region expands indefinitely, the bound-state resonance is delocalized.

\subsection{The $\boldsymbol{\Lambda \to 0^+}$ Limit}

Numerical results in Ref.~\cite{volkel2025} show that the oscillatory region of the eigenfunctions of bound-state resonances expands significantly outward as $n$ increases, and the outer turning point $x_+$ deviates outward exponentially
\begin{equation}
x_+ \approx \sqrt{\frac{l(l+1)}{-E_n}},
\label{eq:79}
\end{equation}
where $E_n \sim -\exp(-K n)$ with $K$ a constant.

To explain the delocalization phenomenon of highly excited bound-state resonances reported in Ref.~\cite{volkel2025}, we focus on the asymptotic behavior of the width of the classically allowed region for highly excited bound-state resonances ($n \gg 1$).

We first analyze modes with $l > 0$, $s \in \{0, \pm1, \pm2\}$ and $l \geq |s|$. From Eq.~\eqref{eq:general_scattering_potential}, we obtain the inverted Regge–Wheeler potential of the asymptotically flat Schwarzschild black hole ($\Lambda=0$). Its asymptotic behavior at large-$|x|$ regime
\begin{equation}
\USch(x; M, s, l) = -\frac{l(l+1)}{x^2} + \calO\left( \frac{\ln x}{x^3} \right) \quad (x \to +\infty),
\label{eq:59}
\end{equation}
and
\begin{equation}
\USch(x; M, s, l) = -\frac{l(l+1)+1-s^2}{4e M^2} e^{x/(2M)} + \calO\left( e^{x/M} \right) \quad (x \to -\infty),
\label{eq:60}
\end{equation}
where $\USch(x; M, s, l)$ is the inverted Regge–Wheeler potential of the Schwarzschild black hole in the tortoise coordinate, and $e$ is the base of the natural logarithm.
Using Eqs.~\eqref{eq:58},\eqref{eq:tortoise_coord},\eqref{eq:general_scattering_potential},\eqref{eq:define inverted potential} and $\Lambda=0$ ,we obtain the coordinates of the classical turning points for highly excited bound-state resonances:
\begin{equation}
x_+(n) = \sqrt{\frac{l(l+1)}{-E_n}} + M \ln\left( \frac{l(l+1)}{-4 M^2 E_n} \right) + \frac{M \left[(1-s^2)-l(l+1)\right]}{l(l+1)} + \calO\left( \sqrt{-E_n} \right),
\label{eq:61}
\end{equation}
and
\begin{equation}
x_-(n) = 2M \ln\left( \frac{-4e M^2 E_n}{l(l+1)+(1-s^2)} \right) + \calO\left( -E_n \right).
\label{eq:62}
\end{equation}

Combining Eqs.~\eqref{eq:61}, \eqref{eq:62} and \eqref{eq:lambda_zero_energy_spectrum}, the width of the classically allowed region is given by
\begin{equation}
\begin{split}
x_+(n) - x_-(n) & = \sqrt{\frac{l(l+1)}{A_{p\ell}}}\times \exp\left( \frac{\pi n}{\sqrt{l(l+1)-1/4}} \right)+ \frac{6\pi M n}{\sqrt{l(l+1)-1/4}} + \calO(1)\\
&\sim  \sqrt{\frac{l(l+1)}{A_{p\ell}}}\times \exp\left( \frac{\pi n}{\sqrt{l(l+1)-1/4}} \right),   
\end{split}
\label{eq:63}
\end{equation}
where
\begin{equation}
A_{p\ell} = \frac{1}{16 M^2} \left\{ \left[ \frac{\Gamma(2\ell+1)}{\Gamma(-2\ell-1)} \right]^2 \frac{\Gamma(-\ell-p)^2 \Gamma(-\ell)}{\Gamma(\ell-p+1)^2 \Gamma(\ell+1)} \right\}^{\frac{1}{\ell+1/2}}.
\label{eq:64}
\end{equation}

Equation \eqref{eq:lambda_zero_energy_spectrum} can be equivalently written as
\begin{equation}
E(p,\ell;n) = - A_{p\ell} \times \exp\left( -\frac{2\pi n}{\sqrt{l(l+1)-1/4}} \right); \quad n \in \mathbb{Z}.
\label{eq:65}
\end{equation}
It follows from Eq.~\eqref{eq:63} that the width of the classically allowed region grows exponentially with increasing $n$. In particular, for highly excited states ($n \gg 1$), the width of the classically allowed region increases rapidly, leading to strong delocalization of the bound-state resonances. This is consistent with the phenomenon observed in the numerical simulations of Ref.~\cite{volkel2025}, where the oscillatory region of the eigenfunction expands significantly outward and the outer turning point deviates outward exponentially. The zeroth-order approximation of Eq.~\eqref{eq:61} is in full agreement with Eq.~\eqref{eq:79}.

The delocalization of bound-state resonances in the inverted potential of asymptotically flat Schwarzschild black holes also explains the phenomenon found in the numerical simulations of Ref.~\cite{volkel2025} that higher energy levels are more sensitive to far-field potential perturbations than lower ones. For a given $x_+(n)$, consider a far-field potential perturbation at $x_0$ satisfying
\begin{equation}
x_+(n) < x_0,
\label{eq:66}
\end{equation}
The perturbing potential lies in the evanescent (classically forbidden) region, making the bound-state resonance relatively insensitive to the perturbation.

For higher energy levels, however, Eqs.~\eqref{eq:61} and \eqref{eq:65} imply that there exists a sufficiently large $n$ such that
\begin{equation}
x_+(n) > x_0.
\label{eq:67}
\end{equation}
This means the perturbing potential falls within the oscillatory (classically allowed) region, rendering the highly excited states sensitive to far-field potential perturbations and thus less localized.

The numerical simulations in Ref.~\cite{volkel2025} mainly focused on asymptotically flat Schwarzschild black holes with $s=\pm2$, $l=2,3,4$, and observed the delocalization of highly excited bound-state resonances and the sensitivity of the wave function to far-field potential perturbations. In this section, we have analytically shown that this phenomenon exists in all modes with $l > 0$, $s \in \{0, \pm1, \pm2\}$ and $l \geq |s|$. For the $l=0$, $s=0$ mode, there exists only one resonant eigenstate for black hole bound-state resonances \cite{LiuMashhoon1996}, and no delocalization occurs.

\subsection{Schwarzschild–de Sitter Black Holes}

In this section, we discuss the corresponding properties of the bound-state resonance system in SdS black holes. For SdS black holes with modes satisfying $s \in \{0, \pm1, \pm2\}$ and $l \geq |s|$, we define the surface gravities
\begin{equation}
\kb = \frac{f'(r_h)}{2} > 0, \quad \kc = -\frac{f'(r_c)}{2} > 0.
\label{eq:68}
\end{equation}

The asymptotic behavior of the inverted SdS potential at large-$|x|$ regime is given by
\begin{equation}
\USdS(x) = -2\kb \left[ \frac{l(l+1)}{r_h^2} + (1-s^2)\left( \frac{2M}{r_h^3} - \frac{(4-s^2)\Lambda}{6} \right) \right]e^{-2\kb H_b} e^{2\kb x} + \calO\left( e^{4\kb x} \right), \quad x \to -\infty,
\label{eq:69}
\end{equation}
where
\begin{equation}
x = \frac{1}{2\kb} \ln\left(r - r_h\right) + H_b + \calO\left(r - r_h\right),
\label{eq:70}
\end{equation}
and $H_b$ is an integration constant. Here $\USdS(x; M, s, l)$ denotes the inverted Regge–Wheeler potential of the SdS black hole in the tortoise coordinate. Similarly,
\begin{equation}
\USdS(x) = -2\kc \left[ \frac{l(l+1)}{r_c^2} + (1-s^2)\left( \frac{2M}{r_c^3} - \frac{(4-s^2)\Lambda}{6} \right) \right] e^{2\kc H_c} e^{-2\kc x} + \calO\left( e^{-4\kc x} \right), \quad x \to +\infty,
\label{eq:71}
\end{equation}
where
\begin{equation}
x = -\frac{1}{2\kc} \ln\left(r_c - r\right) + H_c + \calO\left(r_c - r\right),
\label{eq:72}
\end{equation}
with $H_c$ an integration constant.

Using Eqs.~\eqref{eq:58}, \eqref{eq:tortoise_coord}, \eqref{eq:general_scattering_potential} and \eqref{eq:define inverted potential} , we obtain the coordinates of the classical turning points:
\begin{equation}
x_-(n) = \frac{1}{2\kb} \ln\left(-E_n\right) + \calO(1) = -\frac{1}{2\kb} \ln\left( \frac{1}{-E_n} \right) + \calO(1),
\label{eq:73}
\end{equation}
and
\begin{equation}
x_+(n) = \frac{1}{2\kc} \ln\left( \frac{1}{-E_n} \right) + \calO(1).
\label{eq:74}
\end{equation}

From Eqs.~\eqref{eq:73} and \eqref{eq:74}, the width of the classically allowed region reads
\begin{equation}
x_+(n) - x_-(n) = \left( \frac{1}{2\kb} + \frac{1}{2\kc} \right) \ln\left( \frac{1}{-E_n} \right) + \calO(1).
\label{eq:75}
\end{equation}
It is obvious that the width of the oscillatory region grows logarithmically with increasing energy. Comparing Eqs.~\eqref{eq:63} and \eqref{eq:75}, it is clear that the exponentially rapid delocalization of bound-state resonances seen in asymptotically flat Schwarzschild black holes does not occur for bound-state resonances in SdS black holes.

\subsection{Number of Bound-State Resonance Energy Levels}

From Eqs.~\eqref{eq:69} and \eqref{eq:71}, we have
\begin{equation}
\int_{-\infty}^{-R} (1+|x|) e^{2\kb x} \diff x < \infty,
\label{eq:76}
\end{equation}
and
\begin{equation}
\int_{R}^{\infty} (1+x) e^{-2\kc x} \diff x < \infty,
\label{eq:77}
\end{equation}
where $R$ is a sufficiently large positive constant.

Combining Eqs.~\eqref{eq:69} and \eqref{eq:71}, \eqref{eq:76} and \eqref{eq:77}, we obtain
\begin{equation}
\int_{-\infty}^{\infty} (1+|x|) |\USdS(x)| \diff x < \infty.
\label{eq:78}
\end{equation}
Equation \eqref{eq:78} implies that the number of negative eigenvalues corresponding to bound-state resonance wave functions of $\USdS(x)$ is finite \cite{Faddeev1959}. In other words, the inverted potential admits only a finite number of bound-state resonance energy levels. The bound-state resonances of the asymptotically flat Schwarzschild black hole have an infinite number of bound states \cite{hod2025} except for the $l=0$, $s=0$ mode \cite{LiuMashhoon1996}, which allows the oscillatory region of its eigenfunctions to delocalize indefinitely. In contrast, for SdS black holes in all modes with $s \in \{0, \pm1, \pm2\},l \geq |s|$, the number of bound-state resonance energy levels is finite, which implies that there exists an upper bound on the oscillatory domain of the bound-state resonance eigenfunctions, thus precluding unbounded delocalization. Comparing Eq. \eqref{eq:59} with Eq. \eqref{eq:71}, we find that when $x \to +\infty$, the inverted potential function of the asymptotically flat Schwarzschild black hole decays in an inverse-square manner as  $x$  increases, while that of the SdS black hole decays exponentially. The presence of the cosmological constant profoundly alters the asymptotic behavior of the potential functions, which is the fundamental origin of the structural difference between the bound-state resonance energy spectra of asymptotically flat Schwarzschild black hole and SdS black hole.

\section{DELOCALIZED HALF-BOUND STATE}
By solving the characteristic equation \eqref{eq:sds_characteristic_eq}, we find that there exists an interesting and special solution in the inverted potential of the SdS black hole: the zero-energy resonance. A zero-energy resonance is a half-bound state whose energy level lies exactly at $E=0$ (corresponding to frequency $\omega=0$), and it serves as the critical phase transition point for the system to evolve from``having no bound-states" to ``possessing stable bound-states". Importantly, this half-bound state is delocalized because it is not fully confined by the potential well boundaries. Such delocalized half-bound state exist only in SdS black holes with the cosmological constant taking specific discrete values $\Lambda_m$ $(m\in\mathbb{Z})$, but are absent in asymptotically flat Schwarzschild black holes. 

The analysis in this section is based on electromagnetic perturbation mode ($s=\pm1$).

\subsection{Solution for $\Lambda_m$}
Assume that a zero-energy resonance solution exists when the cosmological constant takes the value $\Lambda_m$. Substituting $k = 0$ ($|\omega|=0$) into the characteristic equation \eqref{eq:sds_characteristic_eq} and solving the equation, we obtain the discrete solutions
\begin{equation}
\Lambda_m= \frac{3}{16M^2} \left[\frac{\Gamma(2\ell+1)\Gamma(-\ell-p) \Gamma(-\ell)}{\Gamma(-2\ell-1)\Gamma(\ell-p+1) \Gamma(\ell+1)} \right]^{\frac{2}{\ell+1/2}} \times e^{-\frac{2\pi m}{\sqrt{l(l+1)-1/4}}}, \quad m\in\mathbb{Z}.
	\label{eq:Solution for lambda_m}
\end{equation}
From Eq. \eqref{eq:Solution for lambda_m}, it follows that $\Lambda_m \not=0$. This means that asymptotically flat Schwarzschild black holes do not possess zero-energy resonance.

\subsection{Critical Point for the Change in the Number of Bound-State Resonances}

We expand the characteristic equation \eqref{eq:sds_characteristic_eq} around $\Lambda=\Lambda_m$, $|\omega|=0$ to obtain
\begin{equation}
 d\Lambda = \frac{\Lambda_m \pi (4M + a) \cot(\pi \ell) d|\omega|}{\ell + \frac{1}{2}},
	\label{eq:Critical Point for the Change}
\end{equation}
Considering Eqs. \eqref{eq:parameter_definition} and  \eqref{eq:Critical Point for the Change}, we find that  $d\Lambda<0$, where $|d\Lambda|\ll|\Lambda_{m+1}-\Lambda_{m}|$.     
As the value of the cosmological constant is decreased from $\Lambda_m$ to $\Lambda_m+d\Lambda$, the half-bound state ($E=0$) transforms into a bound-state resonance ($E=-(d|\omega|)^2$). If the system has $N$ bound-state resonances and one half-bound state when $\Lambda=\Lambda_m$, then it will have $N+1$ bound-state resonances and no half-bound state when $\Lambda\in(\Lambda_m,\Lambda_{m+1})$. 
By extension, the system will have $N+1$  bound-state resonances and one half-bound state when $\Lambda=\Lambda_{m+1}$, and  $N+2$  bound-state resonances and no half-bound state when  $\Lambda\in(\Lambda_{m+1},\Lambda_{m+2})$ (see Table \ref{N,N+1,N+2}). As the value of $\Lambda$ decreases continuously, the system adds one bound-state resonance energy level every time $\Lambda$ crosses the critical phase transition point $\Lambda_m$ . 
In other words, $\Lambda_m$ are the critical phase transition points at which the number of bound-state resonances changes.

\begin{table}[H]
	\centering
	\caption{The correspondence between the values or ranges of $\Lambda$ and the corresponding numbers of bound-state resonances and half-bound state in the system.}
	\label{N,N+1,N+2}
	\begin{tabular}{ccccccccccccc}
		\hline
	 $\Lambda$& ...&($\Lambda_{m-1},\Lambda_{m}$)    &$\Lambda_{m}$   & ($\Lambda_m,\Lambda_{m+1}$)  &$\Lambda_{m+1}$    &($\Lambda_{m+1},\Lambda_{m+2}$) &$\Lambda_{m+2}$  & ...&$\Lambda_{2m}$ &($\Lambda_{2m},\Lambda_{2m+1}$) & ...\\
		\hline
		\# of half-bound state&...  & 0&1  &  0&1 & 0 &1&  ...& 1 &0 &  ...\\
		\# of bound-state resonances& ... & $N$&$N$&$N+1$ &$N+1$ &$N+2 $ &$ N+2$&  ...&$ N +m$&$N+m+1$&...\\
		\hline
	\end{tabular}
\end{table}

\subsection{Relationship between the Number of Bound State Resonances and $\Lambda$}
If the number of bound-state resonances is $N_1$ when the cosmological constant is $\Lambda_A$, and $N_2$ when it is $\Lambda_B$. As the value of cosmological constant decreases from $\Lambda_A$ to  $\Lambda_B$, it will pass through $m$ critical phase transition points $\Lambda_m$, and we obtain 
\begin{equation}
m=N_2-N_1.
	\label{eq:n and N}
\end{equation}
Considering Eq. \eqref{eq:Solution for lambda_m}, we arrive at
\begin{equation}
N_2=N_1+\left\lceil\mu\ln \frac{\Lambda_B}{C_{p\ell}} \right\rceil-\left\lceil\ \mu\ln \frac{\Lambda_A}{C_{p\ell}}  \right\rceil,\ \ \ \ \ (\Lambda_A > \Lambda_B),
\label{N_2-N_1} 
\end{equation}
where
\begin{equation}
	\mu= -\frac{\sqrt{l(l+1)-1/4}}{2\pi }, \quad 
C_{p\ell}= \frac{3}{16M^2} \left[\frac{\Gamma(2\ell+1)\Gamma(-\ell-p) \Gamma(-\ell)}{\Gamma(-2\ell-1)\Gamma(\ell-p+1) \Gamma(\ell+1)} \right]^{\frac{2}{\ell+1/2}} .
\end{equation}
Derived from Eq. \eqref{N_2-N_1}, when $\Lambda_B\to0$, $N_2\to\infty$.
This successfully explains why asymptotically flat Schwarzschild black holes have an infinite number of bound-state resonance energy levels.

\section{CONCLUSIONS}
In this work, we have analytically solved the Schrödinger-like equation (Eq. \eqref{eq:schrodinger_inverted_potential}) governing electromagnetic perturbations ($s=\pm1$) in Schwarzschild–de Sitter (SdS) black holes, working within the regimes of small cosmological constant (Eq. \eqref{eq:small_lambda_inequality}) and low energy (Eq. \eqref{eq:small_energy_inequality}). From this analysis, we derived the characteristic equation (Eq. \eqref{eq:sds_characteristic_eq}) for bound-state resonances in the inverted SdS potential.

By solving the characteristic equation \eqref{eq:sds_characteristic_eq}, we derive compact closed-form analytical formulas for the excited bound-state resonance energy levels $(n\gg1)$ of SdS black holes in specific spectral interval (see Eq. \eqref{E:{1}}), and the $\Lambda$-induced relative shift $\xi_n$ of the bound-state resonance energy spectrum (see Eq. \eqref{eq:xi_n}). These formulas reveals that the influence of the cosmological constant $\Lambda$ is negligible for relatively low-lying resonances but becomes exponentially increasingly significant for higher excitation levels, and even an arbitrarily small $\Lambda$ will exert a non-negligible effect on specific spectral intervals of high-lying excited states. In the limit \(\Lambda \to 0\), our general energy spectrum (Eq.~\eqref{eq:lambda_zero_energy_spectrum}) reduces exactly to the result for asymptotically flat Schwarzschild black holes reported in recent literature \cite{hod2025}, confirming the consistency of our approach. Extending this analytical framework to scalar (\(s=0\)) and gravitational (\(s=\pm2\)) perturbations remains an important direction for future work.

We have also analytically demonstrated that the rapid delocalization of highly excited bound-state resonances---and the associated sensitivity of their wavefunctions to far-field potential perturbations---is a universal feature of asymptotically flat Schwarzschild black holes for all modes with \(l > 0\), \(s \in \{0, \pm1, \pm2\}\), and \(l \geq |s|\). The single resonant eigenstate in the \(l=0, s=0\) mode \cite{LiuMashhoon1996} constitutes the only exception, where no delocalization occurs. The analytical expression for the width of the eigenfunctions' oscillatory region (Eq.~\eqref{eq:63}) shows that this region expands exponentially with the resonance number \(n\). Since \(n\) can grow indefinitely (Eq.~\eqref{eq:lambda_zero_energy_spectrum}), the system exhibits unbounded delocalization for \(n \gg 1\).

For SdS black holes, however, the situation is fundamentally different. We derived the corresponding width of the oscillatory region (Eq.~\eqref{eq:75}) and proved that the unbounded, exponential delocalization present in the asymptotically flat case is absent. Crucially, we established that for all modes with \(s \in \{0, \pm1, \pm2\}\),  \(l \geq |s|\), the spectrum of bound-state resonance energy levels in SdS black holes is finite. This finiteness imposes a strict upper bound on the spatial extent of the eigenfunctions, thereby preventing unbounded delocalization.

Intriguingly, we discover that SdS black holes exhibit a unique class of half-bound state: zero-energy resonances, and these half-bound state are delocalized. Moreover, these delocalized half-bound state occur only in SdS black holes for specific discrete values of the cosmological constant $\Lambda_m$ (see Eq.\eqref{eq:Solution for lambda_m}), and do not exist at all in asymptotically flat Schwarzschild black holes. 

We find that $\Lambda_m$ are the critical phase transition point at which the number of bound-state resonance energy levels changes, derive the relationship between the Number of Bound State Resonance energy levels and $\Lambda$ (see Eq.\eqref{N_2-N_1}), and thereby explain why asymptotically flat Schwarzschild black holes have an infinite number of bound-state resonance energy levels.

This sharp contrast between infinitely delocalized bound-state resonances in asymptotically flat Schwarzschild black holes and the bounded eigenfunctions in SdS spacetimes , as well as the delocalized half-bound states unique to SdS black holes, highlights a profound impact of the cosmological constant on black hole resonance structure. 

\section{ACKNOWLEDGMENTS}
This work was supported by the National Natural Science Foundation of China with the Grant No. 12375049 and No. 12405059, Key Program of the Natural Science Foundation of Jiangxi
Province under Grant No. 20232ACB201008, and the
Ganpo High-Level Innovative Talent Program.

\appendix
\section{Appendix}
\label{app:appendix}

\subsection{Asymptotic expansion of the Gamma function ratio for $\Lambda\to0^+$}
\label{app:gamma_asymptotic}
When $\Lambda\to0^+$, $a|\omega|\to+\infty$ (see Eq. \eqref{eq:far_region_dimless_var2}). In this regime, we can use the standard asymptotic formula for the ratio of Gamma functions \cite{abramowitz1970}
\begin{equation}
\frac{\Gamma(z+\alpha)}{\Gamma(z+\beta)} \sim z^{\alpha-\beta}, \quad |z|\to\infty,
\label{eq:gamma_asymptotic_formula}
\end{equation}
where
\begin{equation}
z = a|\omega|, \quad \alpha = -\ell, \quad \beta = \ell+1.
\label{eq:gamma_asymptotic_params}
\end{equation}
Substituting Eq. \eqref{eq:gamma_asymptotic_params} into Eq. \eqref{eq:gamma_asymptotic_formula} yields
\begin{equation}
\frac{\Gamma(-\ell+a|\omega|)}{\Gamma(\ell+1+a|\omega|)} \sim (a|\omega|)^{-2\ell-1}.
\label{eq:gamma_ratio_asymptotic}
\end{equation}
Taking cognizance of Eqs. \eqref{eq:gamma_ratio_asymptotic} and \eqref{eq:sds_characteristic_eq} one obtains the resonance expression \eqref{eq:lambda_zero_resonance_eq}.

\subsection{Recurrence relation of the Gamma function}
\label{app:gamma_recurrence}
We consider the recurrence relation of the Gamma function \cite{abramowitz1970}
\begin{equation}
\Gamma(x+1) = x\Gamma(x), \quad \Gamma(x+2) = (x+1)x\Gamma(x).
\label{eq:gamma_recurrence_relation}
\end{equation}
where
\begin{equation}
x = -\ell.
\label{eq:gamma_recurrence_x}
\end{equation}
We obtain
\begin{equation}
\Gamma(-\ell+1) = (-\ell)\Gamma(-\ell) = (-\ell)(-\ell-1)\Gamma(-\ell-1) = \ell(\ell+1)\Gamma(-\ell-1),
\label{eq:gamma_neg_ell_recurrence}
\end{equation}
and
\begin{equation}
\Gamma(\ell+2) = (\ell+1)\ell\Gamma(\ell).
\label{eq:gamma_pos_ell_recurrence}
\end{equation}
Taking cognizance of Eq. \eqref{eq:gamma_neg_ell_recurrence} and \eqref{eq:gamma_pos_ell_recurrence}, we obtain
\begin{equation}
\frac{\Gamma(-\ell-1)\Gamma(-\ell+1)}{\Gamma(\ell)\Gamma(\ell+2)} = \frac{\Gamma(-\ell-1)\, \ell(\ell+1)\Gamma(-\ell-1)}{\Gamma(\ell)\, \ell(\ell+1)\Gamma(\ell)} = \frac{\Gamma(-\ell-1)^2}{\Gamma(\ell)^2}.
\label{eq:gamma_identity_proof}
\end{equation}


\end{document}